# A Survey of Visuo-Haptic Simulation in Surgical Training

Felix G. Hamza-Lup[*]
Computer Science
Armstrong Atlantic State Univ.
Savannah, GA, USA

Crenguta M. Bogdan, Dorin M. Popovici
Mathematics and Informatics
Ovidius University
Constanta, Romania

Ovidiu D. Costea
General Surgery
Ovidius University
Constanta, Romania

*Abstract* — **Surgeons must accomplish complex technical and intellectual tasks that can generate unexpected and serious challenges with little or no room for error. In the last decade, computer simulations have played an increasing role in surgical training, pre-operative planning, and biomedical research. Specifically, visuo-haptic simulations have been the focus of research to develop advanced e-Learning systems facilitating surgical training. The cost of haptic hardware was reduced through mass scale production and as haptics gained popularity in the gaming industry. Visuo-haptic simulations combine the tactile sense with visual information and provide training scenarios with a high degree of reality. For surgical training, such scenarios can be used as ways to gain, improve, and assess resident and expert surgeons' skills and knowledge.**

## I. INTRODUCTION

Out of out five known senses, touch is the most proficient. Touch is the only sense capable of simultaneous input and output. Haptics (i.e. haptic technology) is a development of the last two decades that allows the integration of tactile feedback in computer simulations. Visuo-haptic applications are multimodal, allowing the user to receive tactile feedback based on the real properties of simulated objects.

Haptic technology can be applied in a variety of fields but is specifically successful in the gaming industry [1], adding to the entertainment capabilities of existing gaming systems and enriching the user's experience. Another field showing potential for the use of haptics is medical training. The sharp realism needed for effective surgical training, with little or no room for error, makes haptic-based simulators particularly attractive.

Surgical education requires extensive practice on patients with close faculty supervision, and can become cost-ineffective for teaching. Surgical training for specific procedures is often done on animals or cadavers. The Physicians Committee for Responsible Medicine found in a survey of 198 Advanced Trauma Life Support courses nationwide, that more than 90% use human cadavers or simulator dummies for training. The remaining courses use live animals to teach these skills. A critical look at using animals for medical training [2] emphasizes the problems with this approach. The replacement of animal testing and animal experimentation with virtual techniques often yields both ethical and technical advantages. Here is the point where visuo-haptic simulations come into focus. Recent applications of haptic technology include training for simple procedures in dental surgery, or complex procedures for surgical training. Again we emphasize that the rationale for such simulators is also coupled with improvements of ethical and financial nature (i.e. eliminating the need and costs of keeping corpses or live animals for surgical training).

In this paper we provide a survey of the application of the *haptic* paradigm in the medical field, specifically in the training and assessment of resident and novice surgeons. We provide a brief survey of existing technology, APIs, and frameworks, and describe the potential of haptics in surgical training.

The paper is organized as follows: Section 2 contains a review of haptic device characteristics. In Section 3 we provide a brief survey of existing technology for laparoscopic surgical training. Section 4 consists of a brief review of existing APIs and frameworks for the integration of haptics and associated algorithms into interactive simulations. In Section 5, we focus on surgical tasks, the skills necessary for their correct execution, and the existing frameworks for skills assessment. We conclude with a discussion on challenges for development and integration of such simulators in a real hospital environment.

## II. HAPTIC DEVICE CHARACTERISTICS

Haptic research originates with the work of Heinrich Weber, a 19[th] century professor at the University of Leipzig, however robotics was almost non-existent at that time. A few decades later, Lederman and Klatzky [3], summarized four basic procedures for haptic exploration, each bringing forth a different set of object characteristics. The first one, *lateral motion (i.e. stroking)*, provides information about the surface texture of the object; the second, *pressure*, gives information about the firmness of the material; the third, *contour following*, elicits information on the form of the object; and last but not least, *enclosure*, reflects the volume of the object.

During surgical procedures, tactile exploration improves the surgeon's performance, providing additional information besides visual cues. For example, pressure and force magnitude provide information about physiological preexisting tensions at the organ level, and iatrogenic tensions generated upon the organic structures during diagnostic or therapeutic procedures.

The force applied on the unit surface is directly proportional with the physical resistance of the tissues in diverse physiological and pathological situations. Parenchymatous organs are friable, hence a smaller prehension and/or traction force is necessary in comparison with hollow organs, or organs that pose more resistance at traction/torsion. Blood vessels are fragile structures, and the forces that act on them must be significantly smaller in magnitude than the forces on ligament/bone structures.

---

[*] Felix.Hamza-Lup@armstrong.edu

Returning to the hardware, the haptic devices currently available on the market apply relative small forces on the user (usually on the user's hands and/or fingers) through a complex system of servoengines and mechanical links. There are numerous haptic devices on the market, and their price has dropped significantly over the past few years due to mass production. Among the most popular are Sensable's PHANToM® Omni™ and Desktop™ devices that can apply forces through a mechanical joint in the shape of a stylus. As recent as 2007, Novint, a company founded by the researchers of Sandia National Laboratory, marketed the very first commercial haptic device. Falcon Novint (novint.com) has been released on the market at a very low price in conjunction with computer games in the USA, Asia and Australia. Novint licensed key portions of the technology used in Falcons from Force Dimension (www.forcedimension.com), a leading Swiss developer of high-end haptic devices like the Omega.x or Delta.x family.

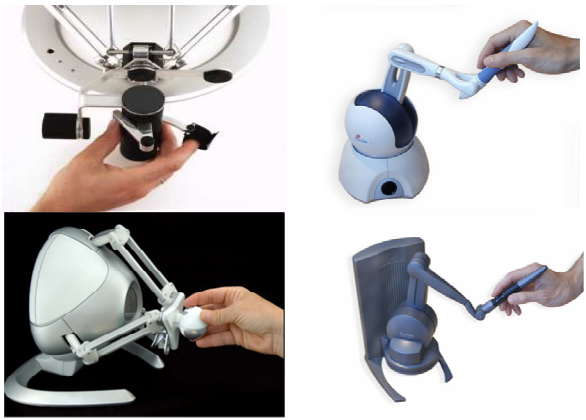

Figure 1. Omega.x (top-left), Falcon Novint™ (bottom-left), PHANToM™ Omni (top-right) and Desktop (bottom-right)

Among the most important characteristics (i.e. *performance measures*) [4] common to all haptic devices, we mention:
- *Degrees-of-freedom*, representing the set of independent displacements that specify the position of the end effectors.
- *Work-volume* refers to the area within which the joints of the device will permit the operator's motion.
- *Position resolution* is the minimum detectable change in position possible within the workspace.
- *Continuous force* is the maximum force that the controller can exert over an extended period of time.
- *Maximum force/torque* is the maximum possible output of the device, determined by such factors as the power of the actuators and the efficiency of any gearing systems. Unlike continuous force, maximum force needs to be exerted only over a short period of time (e.g., a few milliseconds).
- *Maximum stiffness* of virtual surfaces depends on the peak force/torque, but is also related to the dynamic behavior of the device, sensor resolution, and the sampling period of the controlling computer.
- Haptic *update rate* is the inverse of system latency, measured in hertz (Hz).
- *Inertia* is the perceived mass of the device when it is in use. This should be as low as possible to minimize the impact of the device controller on rendered forces.

A novel approach to implementing haptic feedback is through magnetic forces. Magnetic levitation haptic devices allow users to receive force-feedback by manipulating a handle that is levitated within a magnetic field. Users can translate and rotate the handle while feeling forces and torques from the virtual environment. Compared with traditional haptic devices that use motors, linkages, gears, belts, and bearings, magnetic levitation uses a direct electro-dynamic connection to the handle manipulated by the user. Some of the advantages of this approach are: no static friction, no mechanical backlash, high position resolution, simulation of a wide range of stiffness values, and mechanical simplicity. Magnetic haptics has been considered in relation to surgical training systems [5].

The first commercial integration of a magnetic levitation haptic device is the Maglev 200™ Haptic Interface by ButterflyHaptics™ (butterflyhaptics.com), illustrated in Figure 2.

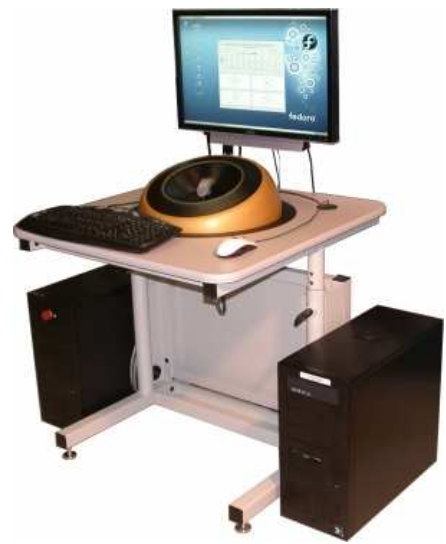

Figure 2. Maglev 200™ Magnetic Levitation Haptic Interface

## III. BRIEF SURVEY OF VISUO-HAPTIC SYSTEMS FOR SURGICAL TRAINING

An early study by Moody et al. [6] demonstrated the effect of a force feedback system in the training and assessment of surgeons. The visuo-haptic system included a PHANToM Desktop unit and simulated a suturing procedure. After the task was demonstrated and explained to each subject by the experimenter, each of the 20 participants performed two test sutures to familiarize themselves with the task and the experimental setting. Participants were then asked to form one suture across a surgical incision, with the specifications provided by the experimenter. Results revealed that force feedback resulted in a reduction of the time taken to complete the stitch.

Most visuo-haptic simulation systems are designed for specific procedures. For example needle insertion is a common procedure that can range in complexity from a simple venipuncture (i.e. to withdraw blood), to a complex procedure such as vertebroplasty (i.e. medical spinal procedure where bone cement is injected through a small hole in the skin into a fractured vertebra with the goal of relieving the pain of osteoporotic compression fractures). Virtual Veins [7] has been primarily used for venipuncture training while a group of researchers at the National University of Singapore developed a surgical simulator for medical student training in the spinal cement vertebroplasty procedure [8]. In vertebroplasty, the surgeon or radiologist relies on sight and feel to properly insert the bone needle through various tissue types and densities. The biomechanical equipment with haptic feedback was designed to capture a user's hand movement and return the tactile information to his fingers allowing him to feel the forces during needle insertion. Other haptic-based simulators involving the task of needle insertion involve spinal injections, [9] and epidural anesthetics (EpiSim - *www.yantric.com*).

From simulation of deformable tissues and their attached properties for the planning of medical procedures [10], to surgical knot-tying procedures [11] and bone surgery [12], visuo-haptic prototypes are being considered. Moreover, when long distance collaboration is necessary, there are prototypes for remote haptic "guidance" of a novice surgeon's hand by an expert surgeon (i.e. *telepresence surgery*) and other applications of Virtual Reality in medicine [13]. Remote training of surgical procedures [14] can improve performance and reduce costs associated with travel.

One of the most promising areas of application for visuo-haptic simulation is laparoscopy (i.e. laparoscopic surgery, non-invasive/minimally invasive surgery) training. Residents as well as experienced surgeons can use these systems for learning, assessing, and improving their surgical procedures and sharpen their skills. The systems have the advantage of changing and adapting the simulation parameters for training under special, unexpected circumstances [15][16].

With the advent of minimally invasive robotic surgery (e.g. daVinci surgical system, www.davincisurgery.com), the haptic coordination of robotic equipment during surgery [17] brought forth new research perspectives. A pneumatic system coupled with sensors at the tip of the tools was proposed to provide haptic feedback to the surgeon during the procedure [18] in a clinical setup.

For training purposes, several companies developed integrated systems that have a set of training scenarios. For example LAP Mentor™ is a multi-disciplinary laparoscopy simulator that enables simultaneous hands-on practice for a single trainee or a team. The system offers training opportunities to residents and experienced surgeons for everything from perfecting basic laparoscopic skills to performing complete laparoscopic surgical procedures.

Another system, the Virtual Endoscopic Surgery Training System One (VSOne), provides force-feedback employing three PHANToM haptic devices and a virtual endoscopic camera. The components are contained in a user-interface box [19] such that they provide an optimal simulated learning environment, similar to a real one. The system contains two applications: VSOne Cho, for laparoscopic cholecystectomy training, and VSOne Gyn, for laparoscopic gynecologic procedures. The following surgical tasks are modeled: grasping, application of clips with coagulation, cutting, irrigation, suction, suturing, and ligation [20]. A series of studies [21] prove that while training with the VSOne system gives similar results as the traditional method, the system reduces the time and cost of training.

A comparable system, CAE Healthcare's (*www.cae.com/en/healthcare*) LapVR surgical simulator, realistically reproduces laparoscopic procedures with haptic technology. The developers claim an accurate simulation of the tactile forces and camera behavior, exactly as it is experienced during laparoscopic surgery.

A survey by Soler et al. [22] claims that the most simulated surgical procedure is the cholecystectomy, available on simulators like LapChole from Xitact (*www.xitact.com*), LapSim from Surgical Science (*www.surgical-science.com*), LapMentor from Simbionix, or RLT from ReachIn (*www.reachin.se*).

While all these development efforts are isolated from each other, and each group developed the systems from basic components and off-the-shelf haptic devices, only a few APIs and frameworks have spawned in recent years. In the next section we provide an overview of the main APIs and frameworks.

## IV. HAPTICS APIS AND FRAMEWORKS

Multiple problems arise in haptic applications interacting with deformable objects. For example, costly computation time, numerical instability in the integration of the body dynamics, time delays etc, may occur. Lengthy computations are forbidden in haptic systems which need high simulation rates (about 1KHz) to obtain realistic force feedback. The update rates of the visual component (i.e. graphic rendering) of the physical objects being simulated is of the order of 20 to 30Hz (frames per second). This difference in the simulation rates can cause an oscillatory behavior in the haptic device that can become highly unstable. Some of these problems can be alleviated with the use of magnetic levitation devices; however, the development of applications in the area is in early research stages.

The most important frameworks and APIs that support the haptic paradigm, and have been used to develop prototypes for commercial applications can be divided into two categories, open source and commercial.

*A. Open Source Frameworks and APIs*

One of the most well known open source API is the Haptics3D (H3D) (*h3d.org*). The API is designed mainly for users who want to develop haptic-based applications from scratch, rather than for those who want to add haptics to existing applications. The main advantages of H3D are the rapid prototyping capability and the compatibility with eXtended 3D (X3D), making it easy for the developer to manage both the 3D graphics and the haptic rendering. For this reason, H3D API is a vital extension to OpenHaptics. It allows users to focus their work on the behavior of the application, and ignore the issues of haptics geometry rendering as well as synchronization of the graphic and the haptic rendering cycles. The API is also extended with scripting capabilities, allowing the user to perform rapid prototyping using the Python scripting language.

Developed with medical applications in mind, the Computer Haptics & Active Interfaces (CHAI) 3D is an open source set of C++ libraries supporting haptic-based systems, visualization, and interactive real-time simulation. The API facilitates the integration of 3D modeling with haptic rendering. Moreover, the applications are portable and can be executed on different platforms. This quality attribute is obtained by saving object characteristics in XML files. The applications can be tested by the programmer using a real haptic device (e.g. PHANToM Omni), or a virtual representation using the mouse as a substitute for the haptic device. The API was recently extended with a simulation engine for rigid and deformable objects.

The need for standardization and inter-project cooperation gave rise to the Simulation Open Framework Architecture (SOFA, *www.sofa-framework.org*). SOFA is targeted at real-time simulation, with an emphasis on medical simulation. The framework allows the development of multiple geometrical models and the simulation of the dynamics of interacting objects using abstract equation solvers. An additional advantage of this framework is the use of the XML standard to streamline the parameters of the simulation like deformable behavior, collision algorithms, and surface constraints.

Another effort targeted at applications of haptics in surgical simulators is the General Physical Simulation Interface (GiPSi) [23]. As the name suggests, GiPSi is a general open source/open architecture framework for developing organ level surgical simulations. The framework provides an API for interfacing dynamic models defined over spatial domains. It is specifically designed to be independent of the specifics of the modeling methods used and therefore facilitates seamless integration of heterogeneous models and processes. The framework contains I/O interfaces for visualization and haptics integration in interactive applications.

*B. Commercial Frameworks and APIs*

The ReachIn API (*www.reachin.se*) is a modern development platform that enables the development of sophisticated haptic 3D applications in the user's programming language of choice, such as C++, Python, or VRML (Virtual Reality Modeling Language). The API provides a base of pre-written code that allows for easy and rapid development of applications that target the specific user's needs. UK Haptics (www.ukhaptics.co.uk), a recently established medical software development company, used ReachIn API as the core haptic technology platform for their Virtual Veins, a medical simulation package for training medical staff in catheter insertion.

To test the flexibility and ease of use of the API, we developed a simple simulation using VRML, Phyton and the ReachIn. In Figure 3, a screenshot of the liver model as seen through the laparoscope camera is illustrated.

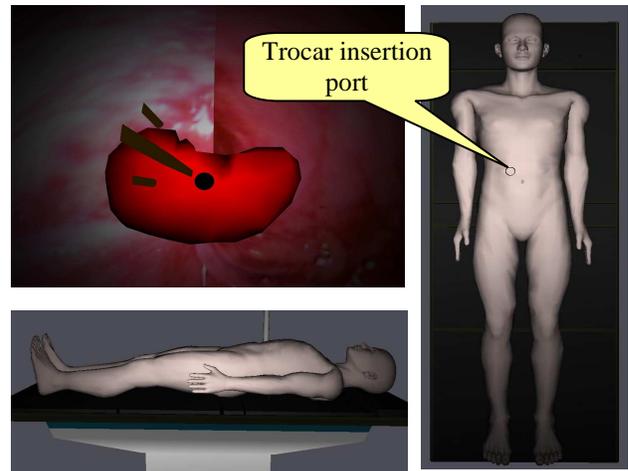

Figure 3. Deformable liver model from SOFA integrated with a humanoid model from MakeHuman (*www.makehuman.org*)

The camera and the light models follow the real laparoscope camera. The haptic feedback is simulated in conjunction with the deformable liver model as well as the humanoid skin surface. The movement of the camera is constrained by the trocar. The light source model follows the camera position and orientation. The conclusion was that ReachIn API is robust and easy to integrate allowing rapid prototype development.

## V. LAPAROSCOPIC SURGICAL PROCEDURES

In the following sections we focus on the surgical tasks and an assessment methodology for interactive visuo-haptic laparoscopy simulators. We present the main skill set and the existing framework for assessment.

### A. Surgical Task Set

Laparoscopic surgical procedures are complex activities that can be decomposed into simple activities called tasks. These tasks can be classified into *basic tasks* and *procedural tasks*. In the laparoscopic cholecystectomy case for example, one encounters the following basic tasks:
- Laparoscope attachments manipulation
- Camera manipulation and navigation
- Light source manipulation and navigation
- Tissue manipulation (e.g. grasping)
- Tissue properties investigation (e.g. soft touch)
- Knot-tying

In the same procedure we encounter the following procedural tasks: suturing, clip application (pre/post resection), surgical cutting, dissecting and separating organs. Some procedural tasks use basic tasks. For example, the suturing task involves knot-tying tasks.

### B. Skill Set

To execute laparoscopic procedures the practitioner must have a series of abilities and skills. For the tasks above, the surgeon must have the following skill set:
- *Basic skills*: e.g. spatio-visual orientation and exploration ability, perceptual abilities, hand-eye coordination, two handed maneuvers, objects relocation.
- *Intermediate skills*: knowing and correctly utilizing the laparoscopic surgery tools for specific cases and the ability to correctly execute the surgical procedure.
- *Advanced skills*: knowledge of the laparoscopic procedures, manual dexterity and precision control.

The above skill classification is based on the performance level of the surgeon and reflects the instruction level (i.e. novice, competent and expert) as well as technical proficiency.

### C. Skill Assessment

Currently the students' skill evaluation is performed by expert surgeons. This makes the evaluation process costly and subjective. Using a visuo-haptic system which supports skill assessment, the subjectivity issues are detached and the probability of human error is reduced.

Since 2001, a taxonomy of metrics for the evaluation of surgical abilities and skills was proposed [24]. This taxonomy is based on two main concepts: *validity* and *reliability*. Each test is designed for a specific objective. The first concept, validity of a test, refers to accepting a test if it is in compliance with five validity measures. The second concept, reliability of a test, refers to the consistency of the results as the test is performed multiple times by the same person or by different persons.

Based on Satava et al. [24], there are five validity measures: face, content, construct, concurrent and predictive. These *validity* metrics endorse the test fulfillment of the objective. Each metric determines the objective fulfillment from a different perspective:
- *Face validity* is determined by the appearance of the interface of the simulated task addressed by the test.
- *Content validity* is determined by the expert surgeons based on the detailed examination of the test content.
- *Construct validity* is determined by the capability of the test to differentiate among performance levels.
- *Concurrent validity* is determined by the capability of the test to return equivalent results with other similar tests.
- *Predictive validity* is determined by the predictive capability of the test, i.e. the evaluated surgeon will have the same performance level in a real scenario.

Two complementary metrics are defined for test *reliability*:
- *Inter-rater reliability*. When the test is performed by two independent evaluators, their results are sufficiently close (if not similar).
- *Test-retest reliability*. Repeating the test at different times and dates should return comparable results.

The validity metric can also be applied in the case of visuo-haptic simulations for laparoscopy procedures. In this case each test is designed for a specific skill, and each validity metric has the following meaning:
- *Face validity*: is determined by the visuo-haptic characteristics of the interface (i.e. how the simulated objects look and feel in comparison with the real objects)
- *Content validity*: if the test measures a certain skill.
- *Construct validity*: the test results should be able to allow differentiation between an expert and a novice surgeon.
- *Concurrent validity*: the capability of a test to return equivalent results with other similar test for the same skill.
- *Predictive validity*: certainty that, after passing the test, the surgeon will have similar performance in a real environment.

Next, we conclude with some of the challenges for the development, and the integration on visuo-haptic simulators in a real hospital environment.

## VI. CONCLUSION

From the development point of view, the APIs and frameworks are currently not interoperable. Even though some effort has been invested recently in developing open frameworks (e.g. GiPSi, SOFA), the software components available are not sufficient to allow rapid development of robust simulation scenarios.

From the integration in a hospital setup perspective, the main challenges are: *budget* - the medical institution/hospital has to allocate funds and faculty "buy-in" time to facilitate the integration of such complex simulators in a clinical setup; *time commitment* - for the faculty, expert surgeons and residents; *suitable space* for setting up training laboratories and required resources. Solutions exist to overcome these challenges from partnerships between industry and education, to employing lower fidelity, inexpensive simulators that can be as effective as expensive simulators for specific tasks.

As a final conclusion, this paper presented a succinct overview of existing visuo-haptic laparoscopic surgical training systems, the existing APIs and frameworks for haptic integration in simulations. We also discussed one of the most important components of visuo-haptic simulators, assessment. We are currently in the process of developing a cost effective battery of visuo-haptic simulation scenarios for laparoscopic surgery. We will report on the progress of this work in future articles.

ACKNOWLEDGEMENTS

This study was supported under the ANCS Grant "HapticMed – Using haptic interfaces in medical applications", no. 128/02.06.2010, ID/SMIS 567/12271 number. We would also like to thank A. Seitan, C. Petre, A. Dinca and M. Polceanu for their contributions.